\documentclass[11pt]{article}
\pdfoutput=1

\usepackage{amstext}
\usepackage{amsfonts}
\usepackage{amssymb}
\usepackage{amsmath}
\usepackage{graphicx}
\usepackage{caption}
\usepackage{subcaption}
\usepackage{float}
\usepackage[pdftex]{hyperref}

\def\ben{\begin{equation}}
\def\een{\end{equation}}
\def\half{{\textstyle{1\over2}}}

\let\a=\alpha   \let\d=\delta

\let\w=\omega \let\G=\Gamma

\def\be{\begin{equation}}
\def\ee{\end{equation}}
\def\beq{\begin{equation}}
\def\eeq{\end{equation}}
\def\ba{\begin{array}}
\def\ea{\end{array}}

\def\dalemb#1#2{{\vbox{\hrule height .#2pt
       \hbox{\vrule width.#2pt height#1pt \kern#1pt
               \vrule width.#2pt}
       \hrule height.#2pt}}}

\newcommand{\bea}{\begin{eqnarray}}
\newcommand{\eea}{\end{eqnarray}}

\def\R{{{\Bbb R}}}

\def\({\left(}
\def\){\right)}

\begin{document}

\begin{center}

{ \LARGE {\bf The Pauli exclusion principle at strong coupling: \\
Holographic matter and momentum space}}

\vspace{1.4cm}

Richard J. Anantua, Sean A. Hartnoll, Victoria L. Martin, David M. Ramirez

\vspace{0.7cm}

{\it  Department of Physics, Stanford University, \\
Stanford, CA 94305-4060, USA }

\end{center}

\vspace{2cm}

\begin{abstract}

For free fermions at finite density, the Pauli exclusion principle is responsible for the existence of a Fermi surface and the consequent
presence of low energy spectral weight over a finite range of momenta. We investigate the extent to which this effect occurs in strongly interacting
quantum matter with a holographic dual. We obtain the low energy current-current spectral weight in two holographic frameworks at finite density: 
systems exhibiting semi-local quantum criticality (with a low temperature entropy density vanishing like $s \sim T^\eta$), and a probe D3/D5 system.
For the semi-local theory with $0 < \eta < 2$ we find a sharp discontinuity in the transverse spectral weight at a nonzero momentum $k_\star$. The case $\eta  = 1$ is found to have additional symmetries and is soluble even at nonzero temperature. We show that this case exhibits a robust linear in temperature resistivity in the presence of random charged impurities.
For the probe D3/D5 system we find an analytic expression for the low energy spectral weight as a function of momentum.
The spectral weight is supported below a specific momentum $k_\star$ and is exponentially suppressed at higher momenta.

\end{abstract}
\thispagestyle{empty}
\pagebreak
\setcounter{page}{1}

\section{Introduction}

The Fermi liquid is a remarkable state of matter. A modern understanding of Fermi liquid theory is as a self-consistently robust organization of the low energy degrees of freedom in metals \cite{Polchinski:1992ed, Shankar:1993pf}. The scaling towards a surface in momentum space can be grounded to some extent in the Luttinger theorem that relates the existence of Fermi surfaces to the fact that there is a charge density \cite{lutt1, lutt2, oshikawa}.
In a simple qualitative picture of a Fermi liquid, the exclusion principle plays a double role.
By preventing the single particle degrees of freedom from condensing in their ground state, it both prevents spontaneous breaking of electromagnetic symmetry (at least at the natural scale set by the chemical potential) and furthermore pushes some low energy excitations of the system out to a finite momentum.

In attempting to move beyond Fermi liquid theory, a crucial question is the extent to which the two roles of the exclusion principle can be decoupled. For instance, perhaps strong interactions persisting to the lowest energy scales can allow the charge-carrying low energy degrees of freedom to spread out over a volume rather than a surface in momentum space, or to collapse to a point in momentum space without triggering Bose condensation. In this paper we will explore this question in two different holographic setups.

The momentum space structure of a Fermi liquid underlies much of its characteristic phenomenology. One should be careful, however. It is sometimes stated that due to having a surface in momentum space of low energy degrees of freedom, a Fermi liquid has infinitely many more degrees of freedom than bosonic systems. The Fermi liquid low temperature specific heat $c \sim T$ is indeed larger than the $c \sim T^d$ specific heat of free relativistic bosons. However, more generally bosons have $c \sim T^{d/z}$, and so there is a quantitative sense in which bosons with dynamical critical exponent $z=d$ have the same number of low energy excitations as a Fermi liquid. Similarly, the logarithmic violation of the area law in the entanglement entropy of Fermi liquids \cite{one,two} is closely tied to the decoupling of low energy degrees of freedom at different points on the Fermi surface \cite{three, four}. However, at least in holographic settings as well as Fermi liquids, this logarithmic violation seems to be tied to the system having a specific hyperscaling violation exponent, $\theta = d - 1$ \cite{Ogawa:2011bz, Huijse:2011ef}. This is a kinematic feature of the system that a priori is logically independent of the momentum space structure of the degrees of freedom \cite{Hartnoll:2012wm}.

Other phenomenological features of Fermi liquids, in contrast, are tied unambiguously to their particular momentum space structure. Quantum oscillations in a magnetic field are a direct consequence of non-analyticity in momentum space \cite{onsager}. The $T^2$ dependence of the resistivity of a clean Fermi liquid due to umklapp scattering depends crucially on the interplay between the lattice and Fermi momenta, e.g. \cite{Hartnoll:2012rj}. Friedel oscillations induced by an external charge depend on the fact that the static susceptibility of free fermions is non-analytic at momentum equal to $2 k_F$, e.g. \cite{wen}.

This paper will focus on the observables that directly characterize the existence of charged low energy excitations as a function of momentum. These are the spectral densities -- imaginary parts of the retarded Green's functions -- of currents and charges. The linear response of a medium to an electromagnetic field may be divided into transverse and longitudinal modes. The quantities of interest are
\be\label{eq:weights}
\sigma_\perp(k) = \lim_{\w \to 0} \frac{\text{Im} \, G^R_{J_\perp J_\perp}(\w,k)}{\w} \,, \qquad  \sigma_\parallel(k) = \lim_{\w \to 0} \frac{\text{Im} \, G^R_{J^t J^t}(\w,k)}{\w}  \,.
\ee
We will compute the spectral densities\footnote{Throughout this paper we will refer to $\text{Im}\, G^R(\w)/\w$ as the spectral density. This quantity is always nonvanishing as $\w \to 0$ at finite temperature, and arises naturally in many contexts.} above in two holographic theories, introduced below. Let us first recall the results obtained for a free nonrelativistic fermion in 2+1 dimensions. Throughout this paper we will consider 2+1 dimensional systems.
The derivation of these formulae is sketched in appendix \ref{sec:free}.
\be
\sigma_\perp(k) =
\left\{ \begin{array}{cc}
{\displaystyle  \frac{\sigma_{0} \sqrt{k_F^2 - k^2/4}}{k}} & k < 2 k_F \\
0 & k > 2 k_F
\end{array} \right.  \,, \quad
\sigma_\parallel(k) =
\left\{ \begin{array}{cc}
{\displaystyle \frac{\sigma_0}{k \sqrt{k_F^2 - k^2/4}}} & k < 2 k_F \\
0 & k > 2 k_F
\end{array} \right. \,.
\label{eq:freetrans}
\ee
In both channels we see that there are no low energy excitations of the system above twice the Fermi momentum $k > 2 k_F$. In the longitudinal channel there is a strong discontinuity at
$k = 2 k_F$, while in the transverse channel it is the derivative of the spectral weight that is discontinuous.
The weaker discontinuity in the transverse channel has a geometrical origin. At low energies the fermions that contribute to the spectral weight are on the Fermi surface. If we now look at external momentum $2 k_F$, the only low energy fermions that could contribute to the conductivity have momentum parallel to the external $k = 2 k_F$ momentum and so do not contribute to the transverse conductivity. The main objective of this paper is to establish the extent to which a momentum space structure analogous to (\ref{eq:freetrans}) exists in known holographic theories.

Before moving on, we would like to contrast the spectral weights (\ref{eq:weights}) with other physically important quantities, the static susceptibilities. These are defined as
\be\label{eq:screen}
\chi_\perp(k) = \lim_{\w \to 0} G^R_{J_\perp J_\perp}(\w,k) \,, \qquad  \chi_\parallel(k) = \lim_{\w \to 0}  G^R_{J^t J^t}(\w,k)  \,.
\ee
The static susceptibilities determine the screening of static currents and charges by the medium. For our free nonrelativistic fermion these can be easily obtained by performing the integrals in appendix \ref{sec:free}. The susceptibilities exhibit a kink at $k = 2 k_F$, but do not vanish for $k > 2 k_F$. Upon Fourier transforming the screened Maxwell field propagator, this kink leads to spatial Friedel oscillations in the response to a point charge. While in a Fermi liquid the static susceptibility is dominated by the low energy degrees of freedom and exhibits a (weaker) non-analyticity at the same momentum as the spectral weight, in general the
static susceptibility is not a direct probe of low energy degrees of freedom. In fact, from the Kramers-Kronig relations, the static susceptibility is equal to the spectral weight integrated over all energies. We will see explicitly in our first holographic example how the static susceptibility cannot be determined from purely low energy data. The focus in the remainder of this paper shall be on the distribution of low energy spectral weight in momentum space.

Holographic descriptions of (compressible) quantum matter at zero temperature exhibit an emergent scaling symmetry in the far interior of the dual spacetime \cite{Hartnoll:2011fn}. This scaling symmetry is characterized by a dynamical critical exponent $z$ \cite{Kachru:2008yh} as well as a hyperscaling violation exponent $\theta$ \cite{Huijse:2011ef, Charmousis:2010zz, Hartnoll:2011fn}. Because the scaling in all known holographic geometries is towards the origin in momentum space, $\w \sim k^z$, the momentum space structure of the low energy charged degrees of freedom is determined by $z$. In particular, if $z < \infty$ is finite, then the spectral weights (\ref{eq:weights}) at any nonzero $k$ are exponentially suppressed as $\w \to 0$ \cite{Hartnoll:2012wm}. In all of these systems with $z < \infty$, a strongly coupled version of the exclusion principle does not appear to be active.\footnote{The discussion in this paragraph and throughout this paper refers to leading order in the bulk semiclassical expansion, corresponding to leading order at large $N$ in the boundary field theory. This is the limit in which holography provides the greatest theoretical and conceptual simplification.} And yet, these geometries can in general be stable against symmetry-breaking instabilities. These non-Pauli excluding geometries include those with logarithmic entanglement entropy \cite{Ogawa:2011bz, Huijse:2011ef}.

In this paper we will obtain the spectral densities (\ref{eq:weights}) in two holographic frameworks that do exhibit a strong coupling remnant of Pauli exclusion physics. The first concerns theories with $z = \infty$. In these systems, time scales at the IR fixed point, but space does not. It is therefore possible to have significant low energy spectral weight at nonzero momentum \cite{Edalati:2010hk, Edalati:2010pn, Hartnoll:2012wm}. Such theories have rightly been termed semi-local quantum liquids \cite{Iqbal:2011in, Iqbal:2011ae}; they are semi-local because while momenta do not scale, the scaling of correlators with energy is generally momentum-dependent. 
We will see examples of this in section \ref{sec:semi} below. In a truly locally critical theory, the IR scaling would be completely independent of momentum. One of the main results in this paper will be the observation that momentum-dependent operator dimensions can lead to important features in the low energy spectral weight. Semi-local quantum criticality can also lead to strong dissipation in fermionic correlators \cite{Faulkner:2009wj}. Semi-locally critical IR geometries are parameterized by the quantity $\eta = - \theta/z$ which is kept finite and positive as $z \to \infty$. They are conformal to $AdS_2 \times \R^2$. At low temperatures, they have an entropy density scaling like $s \sim T^\eta$. For $\eta > 0$ the entropy density vanishes at zero temperature, avoiding the `entropy catastrophe' of pure $AdS_2 \times \R^2$. On the other hand, geometries with $\eta > 0$ have divergent curvatures in the far IR, possibly requiring additional physics to resolve.

The second set of theories we consider are probe brane constructions at finite density. In these cases the charged degrees of freedom are parametrically diluted by neutral modes. In the bulk this means that the brane degrees of freedom do not backreact on the metric and therefore we do not obtain scaling geometries of the form discussed in the previous paragraph. This important physics thrown out in the probe limit is partially compensated for by the fact that the brane action is nonlinear in the bulk Maxwell field. At finite charge density, due to these nonlinearities, in the far IR the equations describing charge and current fluctuations become independent of momentum. This observation suggests that low energy spectral weight will exist at nonzero momenta, reminiscent of Fermi surfaces or semi-local criticality \cite{Hartnoll:2009ns}. Unlike the scaling geometries of the previous paragraph, however, the IR does not decouple in a simple way and the full spacetime is needed to compute the momentum dependence of the spectral densities (\ref{eq:weights}). This computation will be the topic of section \ref{sec:probe}.

The Pauli exclusion principle is a statement about a quantum mechanical wavefunction written in terms of fermionic degrees of freedom.  We are interested instead in strongly interacting effective low energy descriptions of finite density systems that do not admit a quasiparticle description, fermionic or otherwise. Indeed, it would seem to be the case in holographic frameworks that it is not a well defined question, from a universal IR point to view, to ask if the charge is carried by bosons or fermions. In particular, we should resist, perhaps, contaminating our IR perspective with information such as whether the UV charged degrees of freedom in an intersecting brane model are purely fermionic or not (e.g. D3/D5 versus D2/D6 etc.). In this paper we will find that strongly interacting `IR soup' can nonetheless exhibit a feature characteristic of Pauli exclusion, namely the presence of low energy degrees of freedom at nonzero momenta. One is led to wonder about the extent to which Pauli exclusion may be a special case of more profound and general quantum dynamics.

\section{Semi-local quantum liquids}
\label{sec:semi}

The ubiquity of semi-local quantum criticality in holographic theories at finite density is one of the more interesting facts to emerge from a condensed matter perspective on holography. While the simplest $AdS_2 \times \R^2$ geometry may be pathological due to having a macroscopic ground state entropy density, this is not the case for generic semi-locally critical geometries. It was emphasized in \cite{Hartnoll:2012wm} that theories in which $\eta = -\theta/z > 0$ is kept fixed as $z \to \infty$ have semi-locally critical spectral weights, with an entropy density vanishing like $T^\eta$ at low temperatures. These spacetimes are conformal to $AdS_2 \times \R^2$ and appeared in previous works including \cite{Gubser:2009qt, Iizuka:2011hg, Gouteraux:2011ce, Gubser:2012yb}. We will describe the embedding of semi-local criticality into string theory in a particularly interesting case ($\eta = 1$) in section \ref{sec:etaone} below.

Following \cite{Charmousis:2010zz}, using the notation of \cite{Huijse:2011ef, Hartnoll:2012wm}, our starting point is the Einstein-Maxwell-dilaton theory in 3+1 bulk dimensions
\begin{align}
  {\cal L} &= \frac{1}{2\kappa^2} R - \frac{1}{4e^2} Z(\Phi) F^2 -
  \frac{1}{\kappa^2} (\partial \Phi)^2 - \frac{1}{2\kappa^2 L^2}
  V(\Phi) \,.
\end{align}
The low energy spectral weight at low temperatures can be obtained from the IR near-horizon geometry of near-extremal black holes in these theories. We therefore only need the forms of the potential and gauge-kinetic function to leading order as $\Phi \to \infty$, as the dilaton will diverge in the far IR of these solutions. These asymptotic behaviors are taken to be
\be
  Z(\Phi) = r_0^2 Z_0^2 e^{\alpha \Phi} \,, \qquad V(\Phi) = -r_0^2 V_0^2 e^{-\beta \Phi} \,.
\ee
The factors of $r_0^2$ will be convenient shortly. As presented above, $Z_0$ and $V_0$ have units of inverse length. They will carry the units of momentum in the semi-locally critical theory.
This theory admits a scaling solution where the parameters $\alpha$
and $\beta$ will be related to the dynamical critical exponent $z$ and
hyperscaling violation exponent $\theta$. 
Our interest is in the limit $z\rightarrow \infty$ and
$\theta\rightarrow - \infty$ with $\eta= - \theta/z$ held
fixed. This will be seen to correspond to
\be
\alpha = -\beta = \frac{2}{\sqrt{1+2/\eta}} \,.
\ee

Black hole solutions to these theories may be found explicitly and take the form
\begin{align}
  ds^2 &= \frac{L^2}{r^2} \left( -f(r) dt^2 + g(r) dr^2 + dx^2 +
    dy^2\right),& A &= \frac{eL}{\kappa} h(r) dt,& \Phi &=
  \sqrt{1+\frac{2}{\eta}} \log \frac{r}{r_0}.
\end{align}
Evaluated on the logarithmically running dilaton, the functions of the dilaton in the action become $Z(\Phi) = Z_0^2 r^2$ and $V(\Phi) = - V_0^2 r^2$.
The functions $f$, $g$ and $h$ are given by
\be\label{eq:fgh}
  f(r) = \frac{\chi(r)}{r^{4/\eta}} \,, \qquad g(r) = \frac{g_0}{r^2
    \chi(r)} \,, \qquad h(r) = \frac{h_0}{r^{2(1+1/\eta)}} \,,
\ee
where the constants are 
\be
  g_0 = \frac{4}{V_0^2}\left(1+\frac{1}{\eta}\right)^2 \,, \qquad h_0 =
  \frac{1}{Z_0 \sqrt{1+\eta}} \,,
\ee
and the emblackening factor $\chi$ is
\begin{align}
  \chi(r) &= 1-\left(\frac{r}{r_h}\right)^{2(1+1/\eta)}.
\end{align}
Here $r_h$ is the horizon radius, which determines the
temperature of these black hole solutions to be
\begin{align}
  T &= \left(1+\frac{1}{\eta} \right) \frac{r_h^{-2/\eta}}{2
    \pi\sqrt{g_0}} = \frac{V_0}{4\pi} r_h^{-2/\eta}. \label{eq:temp}
\end{align}
From these expressions, we see that the entropy density $s$ indeed vanishes at low temperatures: $s \propto r_h^{-2} \sim T^\eta$. At zero temperature, $\chi = 1$, the metric transforms covariantly under the scaling $t\to \lambda t$, $r \to \lambda^{\eta/2} r$, $ds \to
\lambda^{-\eta/2} ds$. This is our local criticality, time scales but space does not, together with hyperscaling violation parameterized by $\eta$.

To obtain the current-current correlations functions, we need to perturb the solutions above.
Taking the momentum of the perturbations to be in the $x$ direction, the perturbations will decouple into modes that are even or odd under $y \leftrightarrow - y$. Thus upon perturbing, and picking the gauge $A_r = g_{\mu r} = 0$, we will obtain coupled equations for 
the transverse modes $\{\delta A_y, \delta g_{xy}, \delta g_{yt} \}$ and the longitudinal modes $\{\delta
A_t, \delta A_x, \delta g_{tt}, \delta g_{xt}, \delta g_{xx}, \delta
g_{yy}, \delta \Phi \}$.
The perturbations are taken to have the form of a function of $r$ times $e^{- i \w t + i k x}$.

Due to a residual gauge symmetry, the equations for the perturbations above will be a mix of first and second order differential equations. A set of independent and purely second order differential equations are obtained by introducing gauge-invariant variables. These variables are derived in appendix \ref{sec:gaugeinvar} and can be taken to have the following form. In the transverse channel
\be
  \psi_{\perp,1} = \frac{\kappa}{e L} \delta A_y \,, \qquad \psi_{\perp,2} = \frac{r^2}{L^2} ( \omega \delta g_{xy} + k
  \delta g_{yt}) \,,
\ee
and in the longitudinal channel
\bea
 \psi_{\parallel,1} & = & \frac{\kappa}{e L} \partial_r \delta A_t -
  \frac{\sqrt{1+\eta}}{Z_0 L^2 \eta r^{1-2/\eta} \chi(r)} \delta g_{tt} \,, \label{eq:la} \\
   \psi_{\parallel,2} & = & \frac{1}{L^2} \delta g_{yy} + \frac{\eta \chi(r)}{2r \, L^2} \frac{1}{(1+ \eta) (1+ 2
    \hat{k}^2) - (1+2\eta) \chi(r)} \partial_r\left[ r^2 \left(
      \delta g_{xx}+ \delta g_{yy} \right) \right]  \,, \\
    \psi_{\parallel,3} & = & \frac{\delta \Phi}{\sqrt{1+2/\eta}}+ \frac{r^2}{2L^2 } \delta
  g_{yy} \,. \label{eq:lc}
\eea
We have defined $\hat{k} = \sqrt{1+1/\eta} k/ V_0$. From the perturbed Einstein-Maxwell-dilaton equations of motion we obtain coupled equations for each set of the gauge invariant variables. These equations can be further decoupled to a large extent. In the longitudinal case especially, the equations are somewhat messy -- the details can be found in appendix \ref{sec:perteqns}.

\subsection{Transverse channel}
\label{sec:translocal}

The transverse channel at zero temperature was decoupled in \cite{Hartnoll:2012wm}. Here we extend the decoupling to nonzero temperatures. Taking our cue from
\cite{Edalati:2010hk}, we define
\begin{align}
  \varphi_{\perp,\pm}(r) &= \frac{r^{1/2+1/\eta} \chi^{3/2}(r)}{ r^{4/\eta}
    \omega^2 - k^2 \chi(r)} \left[p_\pm(r) \widetilde \psi_{\perp,1}(r) +
    \psi_{\perp,2}'(r)\right] \,,
\end{align}
where we have rescaled  $\widetilde \psi_{\perp,1} = Z_0 \sqrt{1+\eta} \psi_{\perp,1}$ and 
the functions $p_\pm$ are given in appendix \ref{sec:perteqns}. In terms of these new variables, the equations of motion are decoupled
\begin{align}
  0 &= \varphi_{t,\pm}''(r) + \(\frac{q_\pm \left(\chi(r) \right)
    }{4 \eta^2 r^2 \chi^2(r)} + \frac{4(1+\eta)^2 \left[r^{4/\eta}
        \omega^2 - k^2 \chi(r) \right] }{V_0^2 \eta^2 r^2 \chi^2(r) }
  \) \varphi_{t,\pm}(r) \, , \label{eq:nicetran}
\end{align}
where $q_\pm(\chi)$ is given by
\begin{align}
  q_\pm (\chi) &= 4 (1+\eta)^2 - 8(1+\eta)
    \left(3 - \eta \pm 2\sqrt{1 + 2 \eta \hat{k}^2} \right) \chi -
    15\eta^2 \chi^2 \,. \label{eq:tr}
\end{align}
At nonzero temperatures we are only able to solve these equations in the case $\eta = 1$ (as well as the $AdS_2 \times \R^2$ case $\eta = 0$, solved in \cite{Hartnoll:2012rj}), to be discussed in more detail below.

At $T = 0$ we can quickly recover the results of \cite{Hartnoll:2012wm}. At zero temperature $\chi = 1$ in (\ref{eq:nicetran}, \ref{eq:tr}). The resulting equation is solved in terms of Bessel functions for all $\eta$. The solution satisfying infalling boundary conditions at the horizon ($r \to \infty$) is
\begin{align}
  \varphi_{\perp,\pm}(r) &= \sqrt{r} H_{\nu_{\perp}^\pm}^{(1)}\left(
    \frac{r^{2/\eta} \omega (1+\eta)}{V_0} \right)\,,
\end{align}
where the indices are
\begin{align}
  \nu_{\perp}^\pm &= \frac{1}{2} \sqrt{5 + \eta (2+\eta) + 4(1+\eta)^2
    \left(\frac{k}{V_0}\right)^2 \pm 4(1+\eta) \sqrt{1+2 (1+\eta)
      \left(\frac{k}{V_0}\right)^2}}\,.
\end{align}
The retarded Green's function is now found by expanding the solution near the boundary $r \to 0$ \cite{Hartnoll:2009sz, Hartnoll:2012wm}. From the asymptotic expansion of $H_{\nu}^{(1)}(x)$ for $x \to 0$ we obtain
\begin{align}
  \varphi_{\perp,\pm}(r) &\propto \sqrt{r} \left( r^{-2
      \nu_{\perp}^\pm} + {\cal G}_{\perp,\pm}^R(\omega,k) r^{2 \nu_{\perp}^\pm}
  \right) \,,
\end{align}
with
\begin{align}\label{eq:IRgreen}
  {\cal G}_{\perp,\pm}^R(\omega,k) \; \propto \; \omega^{2 \nu_{\perp}^\pm(k)}\,.
\end{align}
This scaling pertains to both the real and imaginary parts of the Green's function.

We are now in a position to obtain an interesting result for the transverse spectral weight.
We have obtained the Green's function in the IR near-horizon geometry, and have not specified how this
is connected onto an asymptotically $AdS_4$ or other spacetime which would define the UV of the system.
Recall, however, that to leading order at low frequencies, the imaginary part of the full UV Green's function is
given by a sum over the imaginary parts of the IR Green's functions of all operators that overlap with the UV operator
\cite{Donos:2012ra}. In the present context, the exponent $\nu_{\perp}^-$ leads to the strongest IR behavior
and so we obtain that at small frequencies
\be\label{eq:grjj}
\text{Im} \, G^R_{J_\perp J_\perp}(\w,k) \; \propto \; \text{Im} \, {\cal G}_{\perp,-}^R(\omega,k) \; \propto \; \omega^{2 \nu_{\perp}^-(k)} \,.
\ee
And therefore the transverse spectral weight (\ref{eq:weights}) is found to be
\be
\sigma_\perp(k) = \lim_{\w \to 0} \frac{\text{Im} \, G^R_{J_\perp J_\perp}(\w,k)}{\w} =
\left\{ \begin{array}{cc}
\infty & \qquad k < k_\star \\
0 & \qquad k > k_\star
\end{array} \right. \,.\label{eq:infzero}
\ee
The critical momentum is given by
\be\label{eq:kcritlocal}
k_\star^2 = V_0^2 \frac{\eta (2 - \eta) + 2 \sqrt{2 \eta (1+\eta)(2 - \eta)}}{4 (1+\eta)^2} \,, \qquad (0 < \eta < 2) \,.
\ee
Outside the range $0 < \eta < 2$, the transverse spectral weight is zero everywhere.

The reader that finds (\ref{eq:infzero}) unpalatable, or worries that dividing by some given power of $\w$ is artificial, might like to turn on a small temperature. While we have not solved the equations in general at finite temperature, it is straightforward to show that at low temperatures and for $\w \ll T$ one has
\be\label{eq:lowTgen}
\text{Re} \;  {\cal G}^R(\omega,k) \sim T^{2
    \nu} \,, \qquad \text{Im} \;  {\cal G}^R(\omega,k)
  \sim \omega \, T^{2\nu-1} \,.
\ee
This expression is true for all $\eta$ in both the transverse and longitudinal channels. It follows in particular that
\be\label{eq:therm}
\sigma_\perp(k) \; \propto \; T^{2 \nu_{\perp}^-(k) - 1} \,.
\ee
This expression is finite and nonzero for all momenta. As we take the low temperature limit, the spectral weight diverges for $k < k_\star$ and goes to zero for $k > k_\star$.

The result (\ref{eq:infzero}) for the zero temperature transverse spectral weight, or the thermally regularized version (\ref{eq:therm}), dramatically reveals the existence of low energy spectral weight over a nonzero but finite range of momenta. In our theories with $0 < \eta < 2$ -- this does not include the simplest case of $AdS_2 \times \R^2$ -- a strongly interacting cousin of Pauli exclusion appears to be operating. The physics is tied in a precise way to the emergence of semi-local criticality \cite{Iqbal:2011in}. For $k < k_\star$ the leading IR operator excited by the transverse current is relevant under the local scaling, while for $k > k_\star$ it is irrelevant. This follows from the fact that the dimension of the IR operator is $\Delta = \nu + \half$, and so becomes relevant if $\nu < \half$. But $\nu < \half$ corresponds to the range of momenta over which the zero temperature spectral weight diverges in (\ref{eq:infzero}).

One must be careful not to equate the $k=0$ values of (\ref{eq:IRgreen}) and (\ref{eq:therm}) with the optical and d.c. conductivities. For the conductivities, it is crucial that the limit $k \to 0$ be taken before the $\w \to 0$ limit. At exactly $k=0$, the two gauge-invariant transverse modes decouple. The transverse current operator at $k=0$ is found to correspond to the index $\nu_\perp^+(0)$, and not $\nu_\perp^-(0)$ as one would obtain by naively evaluating (\ref{eq:IRgreen}) and (\ref{eq:therm}) at $k=0$. The optical conductivity is therefore $\sigma \sim \w^{2 \nu_{\perp}^+(0) - 1} = \w^{2+\eta}$, while the critical contribution to the d.c. conductivity is $\sigma_\text{d.c.} \sim T^{2+\eta}$. These expressions agree with previous results in e.g. \cite{Iizuka:2011hg}. The critical d.c. conductivity is swamped by a translation-invariance delta function. In section \ref{sec:etaone} below we discuss the effects of impurities that render the d.c. conductivity finite. There we will obtain a strong d.c. resistivity, that is linear in temperature when $\eta = 1$.

The critical momentum $k_\star$ is set in (\ref{eq:kcritlocal}) by the data $\{\eta,V_0\}$ that appear in the definition of the bulk theory and the IR scaling solution. It is natural to consider the dimensionless ratio of the critical momentum divided by the (square root of the) total charge density of the system.
The total charge density is defined in the UV but, in the absence of bulk charges, can also be obtained from the IR because Gauss's law implies that the integrated flux is independent of radius \cite{Hartnoll:2011pp}. The charge density is then
\be
Q = \int_{\R^2} \star \left(Z(\Phi) F^2 \right) = - \frac{Z(\Phi) h'}{\sqrt{f g}} = \frac{V_0 Z_0}{\sqrt{1 + \eta}} \,.
\ee
The dimensionless ratio is therefore
\be
\frac{k_\star^2}{Q} = \frac{V_0}{Z_0} \frac{\eta (2 - \eta) + 2 \sqrt{2 \eta (1+\eta)(2 - \eta)}}{4 (1+\eta)^{3/2}} \,.
\ee
This ratio can be obtained from the parameters characterizing the IR Lagrangian, i.e. it is a property of the IR theory, but does not appear to be a universal number.

We have found the low energy spectral weight (\ref{eq:infzero}) using only IR data about the theory. As we anticipated in the introduction, this is not the case for the static susceptibilities (\ref{eq:screen}). Recall that, schematically, the full zero temperature Green's function at low frequencies takes the form
\be
G^R_{J_\perp J_\perp}(\w,k) \sim \frac{A(k) + B(k) {\cal G}_{\perp,-}^R(\omega,k)}{C(k) + D(k) {\cal G}_{\perp,-}^R(\omega,k) } \,.
\ee
Here $\{A,B,C,D\}$ are real functions of momentum that are determined by solving the bulk differential equations in the `far' region, away from the horizon, at $\w = 0$. The full structure is in fact more complicated than this, due to mixing among different modes between the IR and the UV \cite{Donos:2012ra}, but the above expression is sufficient for illustrative purposes.
Because $\nu_{\perp}^-(k) > 0$ in the IR Green's function (\ref{eq:IRgreen}), when we take $\w \to 0$ in the full Green's function to obtain the susceptibility, the contribution from the IR Green's function will go to zero. We are left with $\chi_\perp(k) \sim A(k)/C(k)$. This function of momentum is not determined by near-horizon IR data, but requires solving the Einstein-Maxwell-dilaton equations in the full geometry. It follows that (\ref{eq:infzero}) cannot be directly related to the existence of (transverse) Friedel oscillations.

\subsection{Longitudinal channel}

The equations of motion describing the three gauge-invariant longitudinal perturbations (\ref{eq:la}) -- (\ref{eq:lc}) are given in appendix \ref{sec:perteqns}. It is possible to obtain decoupled equations for two linear combinations of the three modes, but the resulting equations are still somewhat messy. Here we present results at zero temperature, while below we will solve the equations at finite temperature in the case $\eta = 1$.

For $T=0$, the longitudinal equations given in appendix \ref{sec:perteqns} simplify to
\begin{align}
  0 = \varphi_{\parallel,i}'' + \frac{1}{r^2} M_{ij}\varphi_{\parallel,j} +
 \frac{4(1+\eta)^2}{r^2 \eta^2 V_0^2} (r^{4/\eta}
  \omega^2 -k^2)  \varphi_{\parallel,i} \,.
\end{align}
Here $\varphi_{\parallel,i}$ ($i = 1,2,3$) are linear combinations of the gauge-invariant modes (\ref{eq:la}) -- (\ref{eq:lc}) and are defined in the appendix.
The matrix $M$ has constant entries given by
\begin{align}
  M &=
  \begin{pmatrix}
    -\frac{3}{4} - \frac{4}{2+\eta} - \frac{9+14\eta}{ \eta^2} &
    0 & - \frac{16(1+\eta)^2 \hat{k}^2 }{\eta^2} \\
    -\frac{8 (1+\eta) \left(\hat{k}^2-1\right)}{ \eta \left( 2
        (1+\eta) \hat{k}^2 -\eta \right)} & - \frac{(2+\eta) (2 +
      3\eta)}{4 \eta^2} & \frac{16(1+\eta) \hat{k}^2 }{ \eta \left(2
        (1+\eta) \hat{k}^2 -\eta \right)} \\
    - \frac{4 (1+\eta)}{\eta (2+\eta)} & 0 & - \frac{(2+\eta) (2+
      3\eta)}{4 \eta^2}
  \end{pmatrix} \,.
\end{align}
By diagonalizing $M$ we obtain three decoupled differential
equations. These can be solved in terms of Bessel functions. Imposing
infalling boundary conditions at the horizon $r \to \infty$ gives the solutions again in terms of a Hankel function
\be
  \varphi_{\parallel,a}(r) = \sqrt{r} H_{\nu_{\parallel}^{a}}^{(1)}\left(
    \frac{r^{2/\eta} \omega (1+\eta)}{V_0} \right) \,.
\ee
Here $a= \{0,+,-\}$ labels the decoupled modes
and the indices are given by
\begin{align}
  \nu_{\parallel}^{0}(k) &= \frac{1+\eta}{2} \sqrt{1+4 \left( 
      \frac{k}{V_0}\right)^2} \,, \label{eq:lnu1} \\
  \nu_{\parallel}^{\pm}(k) &= \frac{1+\eta}{2\sqrt{2+\eta}} \sqrt{10 + \eta +
    4 (2+\eta) \left(\frac{k}{V_0} \right)^2 \pm 8 \sqrt{1+(2+\eta)
      \left(\frac{k}{V_0} \right)^2}} \,. \label{eq:lnu2}
\end{align}
Similarly to the transverse channel, from the asymptotic expansion of the Hankel function we find that the Green's functions have scaling
\be
  {\cal G}_{\parallel,a}^{R}(\omega, k) \propto \omega^{2\nu_{\parallel}^a} \,.
 \ee

Unlike in the transverse case, we see in equations (\ref{eq:lnu1}) and (\ref{eq:lnu2}) that for all $\eta$ and momentum, the indices $\nu_\parallel^a \geq \half$. It follows that the longitudinal low energy spectral weight (\ref{eq:weights}) always vanishes. It does not exhibit the dramatic divergence below a critical momentum that we found in (\ref{eq:infzero}) for the transverse sector.
Relatedly, the finite temperature low energy spectral weight (\ref{eq:therm}) will now be uniformly small at low temperatures over all momenta. In terms of semi-local criticality, this is the statement that the longitudinal operators are irrelevant in the IR theory for all momenta.

It is interesting that these systems, with $0 < \eta < 2$, exhibit a sharper structure in momentum space in the transverse channel than in the longitudinal channel. This is in contrast to the free fermions that we reviewed in the introduction. It is perhaps natural that the longitudinal channel operators are irrelevant in the semi-locally critical picture. The charge density at $k=0$ cannot be relevant: that would imply we had not isolated the correct finite density IR fixed point. It seems reasonable, in addition, to expect that the scaling dimension will increase with $k$. The irrelevance of the longitudinal channel operators for all momenta further implies that our semi-locally critical IR theories are robust under addition of an ionic lattice \cite{Hartnoll:2012rj}.

\subsection{The magical $\eta=1$ case: enhanced symmetries and linear resistivity}
\label{sec:etaone}

In this section we will find that the case with $\eta = 1$ has some interesting mathematical and phenomenological properties.
This is perhaps made more interesting by the fact that this case arises easily from string theory compactifications.\footnote{We thank Aristomenis Donos for pointing out to us that these embeddings all realize $\eta = 1$.} Extremal black holes in $AdS_4 \times S^7$ with one of the four $U(1)$ charges being zero all have a near horizon geometry that is locally critical with $\eta=1$ \cite{Cvetic:1999xp}. These critical geometries therefore describe finite density phases of the worldvolume theory of $N$ coincident $M2$ branes \cite{Maldacena:1997re}.
Furthermore, if the three nonvanishing charges are equal, the background solutions can be obtained from a truncated theory with one gauge field and one scalar, of the type we are considering in this paper (e.g. \cite{Gubser:2009qt}, note however that finite momentum perturbations will probably require a larger set of fields due to an $F \wedge F$ term in the action that sources axionic fields \cite{Cvetic:1999xp}).

For $\eta=1$, the transverse equations (\ref{eq:nicetran}) reduce to
\begin{align}
  0&= \varphi_{\perp,\pm}''(r) + \left(\frac{16\left(r^{4}\omega^2 -
        k^2 \chi(r) \right)}{V_0^2 r^2 \chi^2(r)} + \frac{16 - 15
      \chi^2(r)}{4 r^2 \chi^2(r)} - \frac{8\left(1 \pm \sqrt{1+ 2
          \hat{k}^2} \right)}{r^2 \chi(r)} \right)
  \varphi_{\perp,\pm}(r) \,.
\end{align}
This value of $\eta$ is the only nonzero value for which we have been
able to solve the finite temperature equations analytically.  This
equation can be solved in terms of hypergeometric functions. The
solution satisfying infalling boundary conditions at the horizon is
given by 
\begin{align} \varphi_{\perp,\pm}(r) = r^{1/2} \chi(r)^{ \frac{1}{2}-
    \frac{i \omega}{4 \pi T}} \Big( &\G(c) \G(d)\G(1 -
  \nu_{\perp}^\pm)\left(r/r_h\right)^{2\nu_{\perp}^\pm} \; {}_2F_1
  \left(c,d, 1+
    \nu_{\perp}^\pm, r^4/r_h^4 \right) \notag \\
  &+ \; \{\nu_{\perp}^\pm \to - \nu_{\perp}^\pm \} \Big) \,,
\end{align}
where $c = -\frac{i \omega}{4 \pi T} +
\frac{\nu_{\perp}^\pm}{2}$, $d = 1-\frac{i \omega}{4 \pi T} +
\frac{\nu_{\perp}^\pm}{2}$ and the indices are those quoted in section
\ref{sec:translocal}, setting $\eta=1$:
\begin{align}
  \nu_{\perp}^\pm &= \sqrt{2} \sqrt{1+ \hat k^2
    \pm \sqrt{1+ 2 \hat k^2}} \,,
\end{align}
and we used the temperature $T = V_0/(4 \pi r_h^2)$, from (\ref{eq:temp}) with $\eta = 1$.

Expanding near the boundary, as in previous sections, we obtain the
retarded Green's functions
\begin{align}
  {\cal G}_{\perp,\pm}^R(\omega,k) &\propto T^{2\nu_{\perp}^\pm} 
    \frac{\Gamma \left(- \frac{i \omega}{4 \pi T} +
        \frac{\nu_{\perp}^\pm}{2} \right)}{\Gamma \left(- \frac{i
          \omega}{4 \pi T} - \frac{\nu_{\perp}^\pm}{2} \right)} \frac{\Gamma \left( - \frac{i \omega}{4 \pi T} + 1+
        \frac{\nu_{\perp}^\pm}{2} \right)}{\Gamma \left( - \frac{i
          \omega}{4 \pi T} +1- \frac{\nu_{\perp}^\pm}{2} \right)} \,. \label{eq:sl2sl2}
\end{align}
Expanding this expression at small frequencies gives
\be
  \text{Re} \;  {\cal G}_{\perp,\pm}^R(\omega,k) \sim T^{2
    \nu_{\perp}^\pm} \,, \qquad \text{Im} \;  {\cal G}_{\perp,\pm}^R(\omega,k)
  \sim \omega T^{2\nu_{\perp}^\pm-1} \,.
\ee
Here we have recovered explicitly the general result (\ref{eq:lowTgen})
that holds for all values of $\eta$.

The Green's function (\ref{eq:sl2sl2}) is of a form that strongly suggests the existence of hidden $SL(2,\R)$ symmetries. More precisely, it is the form that follows from an $SL(2,\R) \times SL(2,\R)$ symmetric system in which a hyperbolic generator has been `level matched' \cite{Anninos:2011af}. It seems likely that this symmetry is directly related to the $AdS_3$ factor that appears in the string theory uplift of these geometries \cite{Gubser:2009qt}. See \cite{Balasubramanian:2007bs} for a related earlier appearance of $AdS_3$ in these types of solutions. This offers the hope of obtaining an explicit dual to our IR solutions and thereby a microscopic understanding of the hyperscaling violation and the nature of the momentum dependence of the spectral weight.

Turning now to the longitudinal channel, at finite temperatures and with $\eta = 1$,
the equations for $\varphi_{\parallel,i}$ in appendix \ref{sec:perteqns}
simplify to
\begin{align}
 \varphi_{\parallel,i}'' + \left[\frac{M_{ij}}{r^2 \chi(r)} + \frac{N_{ij}}{r^2}
  \right] \varphi_{\parallel,j}  = -\left(\frac{4}{r^2 \chi^2(r)} + \frac{16\left(r^4 \omega^2 - k^2
        \chi(r)\right)}{ V_0^2 r^2 \chi^2(r)}
  \right)  \varphi_{\parallel,i} \,,
\end{align}
where the matrices $M$ and $N$ are given by
\begin{align}
  M &= \begin{pmatrix} -\frac{88}{3} & 0 & - 64 \hat{k}^2 \\ - \frac{4
      (2 \hat{k}^2 -3)}{2 \hat{k}^2 + 1} & 8 & \frac{16 \left(1 - 2
        \hat{k}^2 \right)}{1 + 2 \hat{k}^2} \\ - \frac{8}{3} & 0 & -8
  \end{pmatrix} \,, & N &= \begin{pmatrix} \frac{1}{4} & 0 & 0 \\ -
    \frac{4}{1 + 2 \hat{k}^2} & - \frac{63}{4} & \frac{16 (3 \hat{k}^2
      - 1 )}{2 \hat{k}^2 + 1} \\ 0 & 0 & \frac{1}{4}
  \end{pmatrix}.
\end{align}
Magically, it seems to us, these two matrices commute so they can be simultaneously
diagonalized. The diagonalized equations can then be solved analytically in terms of hypergeometric functions, rather similarly to the transverse case. The infalling solution is
\begin{align}
  \varphi_{\parallel,a}(r) = r^{1/2} \chi(r)^{ \frac{1}{2}- \frac{i
      \omega}{4 \pi T}} \Big( &\G(c) \G(d)\G(1 -
  \nu_{\parallel}^a)\left(r/r_h\right)^{2\nu_{\parallel}^a} \; {}_2F_1
  \left(c,d, 1+ \nu_{\parallel}^a, r^4/r_h^4 \right) \notag \\
  &+ \; \{\nu_{\parallel}^a \to - \nu_{\parallel}^a \} \Big) \,.
\end{align}
Here $c = s_a -\frac{i \omega}{4 \pi T} +
      \frac{1+\nu_{\parallel}^a}{2}$ and $d =  - s_a-\frac{i \omega}{4 \pi T}
      + \frac{1+\nu_{\parallel}^a}{2}$.
As above, $a = \{0,+,-\}$ labels the decoupled modes, and $s_0 = 1$ while $s_\pm = 0$.
The indices are obtained by setting $\eta=1$ in our previous expressions (\ref{eq:lnu1}) and (\ref{eq:lnu2})
\be
  \nu_\parallel^0 = \sqrt{1 + 2 \hat k^2} \,, \qquad \nu_\parallel^\pm =
  \sqrt{\frac{11}{3} + 2 \hat k^2 \pm \frac{8}{3} \sqrt{1+ \frac{3}{2} \hat k^2}} \,.
\ee

Once again imposing ingoing boundary conditions at the horizon and expanding near the asymptotic boundary, we obtain the retarded Green's functions
\begin{align}
 {\cal G}_{\parallel,a}^R(\omega,k) &\propto T^{2\nu_{\parallel}^a} 
    \frac{\Gamma \left(- \frac{i \omega}{4 \pi T} +
        \frac{1 + \nu_{\parallel}^a}{2} + s_a \right)}{\Gamma \left(- \frac{i
          \omega}{4 \pi T} + \frac{1 - \nu_{\parallel}^a}{2} + s_a \right)} \frac{\Gamma \left( - \frac{i \omega}{4 \pi T} +
        \frac{1 + \nu_{\parallel}^a}{2} - s_a \right)}{\Gamma \left( - \frac{i
          \omega}{4 \pi T} + \frac{1 - \nu_{\parallel}^a}{2} - s_a \right)} \,.\label{eq:sl2again} \end{align}
These are again found to have the form expected from an underlying level matched $SL(2,\R) \times SL(2,\R)$ symmetry. The transverse Green's function (\ref{eq:sl2sl2}) is also in the form
of the above equation (\ref{eq:sl2again}) with $s_\perp = \half$. Comparing these Green's functions with those derived from an $SL(2,\R)\times SL(2,\R)$ symmetry in \cite{Anninos:2011af}, we see that the modes live in a symmetric lowest weight representation of $SL(2,\R) \times SL(2,\R)$ with weight $\Delta = (\nu + 1)/2$
and the matched hyperbolic (i.e. $H-K$ generator in the $SL(2,\R)$ algebra) momentum is, in the notation of \cite{Anninos:2011af}, $x = 2 s$.

In possession of the finite temperature longitudinal spectral weight, we can obtain an interesting phenomenological result. In particular we consider the d.c. electrical resistivity. To obtain a nonzero resistivity it is crucial to break translation invariance. One way to do this is via random impurities. For a locally critical theory in the presence of random charged impurities, the density-density spectral weight (\ref{eq:weights}) controls the d.c. electrical resistivity through the formula 
\cite{Hartnoll:2012rj, Hartnoll:2008hs, Hartnoll:2007ih}
\be
r_\text{imp.} \sim \int \frac{d^2k}{(2\pi)^2} k^2 \lim_{\w \to 0} \frac{\text{Im} \, G^R_{J^t J^t}(\w,k)}{\w} \sim \int dk k^3 T^{2 \nu_\parallel^-(k) - 1}\,.
\ee
The function $\nu_\parallel^-(k)$ is monotonically increasing with $k$. At low temperatures the integral is therefore dominated by small momenta. Up to corrections that are logarithmic in temperature, the leading low temperature behavior of the resistivity is found to be
\be\label{eq:imp}
r_\text{imp.} \sim T^{2 \nu_\parallel^-(0) - 1} = T^{\eta} \,.
\ee
Curiously, we see that precisely the value $\eta = 1$, which we have just seen to have simple string theory realizations and enhanced symmetries relative to the other cases, gives a linear in temperature resistivity. A linear in temperature resistivity is widely observed in quantum critical metals, see e.g. \cite{Sachdev:2011cs}.

The density-density spectral weight also controls the d.c. resistivity due to umklapp scattering by an ionic lattice. It was shown in \cite{Hartnoll:2012rj} that in a locally critical theory, umklapp scattering leads to a resistivity
\be
r_\text{umklapp} \sim T^{2 \nu_\parallel^-(k_L) - 1} \,,
\ee
where $k_L$ is the lattice momentum. This result is UV sensitive, as the power itself depends on lattice data. In contrast, the impurity resistivity (\ref{eq:imp}) only depends on the IR theory.
However, it is worth noting\footnote{We thank Diego Hofman for emphasizing this point to us in discussions.} that the functions $\nu_\parallel^-(k)$ are quite flat at small momenta, with the correction to the $k=0$ value going like $k^4$. It follows that there will be a range of lattice momenta for which the d.c. resistivity due to umklapp scattering will also be very close to linear in temperature when $\eta  =1$.

\section{The D3/D5 system at finite density}
\label{sec:probe}

In this section we obtain the momentum space distribution of the low energy degrees of freedom in a finite density holographic system with a known supersymmetric field theory dual. The field theory will be ${\mathcal N} = 4$ $SU(N)$ SYM theory at large $N$ coupled to a single massless hypermultiplet in the fundamental representation. The hypermultiplet is furthermore constrained to propagate on a 2+1 defect in the full 3+1 field theory dimensions. This system is known to be holographically dual to a single probe $D5$ brane wrapping $AdS_4 \times S^2$ in the $AdS_5 \times S^5$ background of type IIB string theory \cite{Karch:2000gx, DeWolfe:2001pq, Erdmenger:2002ex}. We will consider the system at finite density under the `baryonic' global $U(1)$ symmetry that is dual to the Maxwell symmetry of the bulk probe $D5$ brane. There is a large literature considering Dp/Dq systems in the probe limit at finite density, a slower exposition of the setup we consider may be found in \cite{Goykhman:2012vy}. Our results will hold for many other brane systems beyond the explicit supersymmetric D3/D5 theory we have chosen to study.

The background spacetime is the Schwarzschild black hole in $AdS_5 \times S^5$
\be
ds^2 = L^2 \left(\frac{1}{r^2} \left(- f(r) dt^2 + \frac{dr^2}{f(r)} + dx^2 + dy^2 + dz^2 \right) + d\Omega^2_{S^5} \right) \,.
\ee
Here the emblackening factor $f(r) = 1 - r^4/r_+^4$. The temperature of the black hole is given by $ T = 1/\pi r_+$, see e.g. \cite{Hartnoll:2009sz}.
The probe $D5$ brane is described by the DBI action
\be
S_{D5} = -  e^{-\phi} T_{D5} \int d^6\sigma \sqrt{- \text{det} \left({}^\star g + 2 \pi \a' F \right)} \,.
\ee
Here ${}^\star g$ is the pullback of the metric to the worldvolume, $F = dA$ is the field strength of the worldvolume Maxwell field. The prefactors are the string coupling and the $D5$ brane tension.
For the configurations we consider, the Wess-Zumino term in the brane action, and corresponding equations of motion, is zero. Taking the embedding $AdS_4 \times S^2$ in the background spacetime, corresponding to massless fundamental fields, all that is necessary to place the system at finite density is to solve for the electrostatic potential $A_t(r)$ on the worldvolume. The equations of motion are easily solved to give
\be\label{eq:stat}
A_t'(r) = - \frac{Q}{\sqrt{1 + \hat Q^2 r^4}} \,.
\ee
Here we have introduced $\hat Q^2 = \frac{4 \pi^2}{\lambda} Q^2$, where the 't Hooft coupling
$\lambda \gg 1$ appears through $L^4 = \lambda \a'^2$.
The constant of integration $Q$ is proportional to the charge density of the system. This can be seen by expanding (\ref{eq:stat}) near the boundary at $r=0$. More convenient for us will be the chemical potential
\be\label{eq:mu}
\mu = - \int_0^{r_+} dr A_t'(r) = \frac{Q}{\pi T} \, {}_2 F_1 \left(\frac{1}{4},\frac{1}{2},\frac{5}{4}; - \frac{\hat Q^2}{(\pi T)^4}\right) \,.
\ee

In the probe limit, the excitations of the Maxwell field do not backreact on metric modes. This is technically a great simplification compared to the case we considered in the previous section.
Only the following perturbations are necessary: $\d A_t, \d A_x, \d A_y$. Taking each mode to be a function of $r$ times $e^{- i \w t + i k y}$, one finds the equation for the transverse mode
\be\label{eq:perp}
\d A_\perp'' + \left(\frac{f'}{f} + \frac{2 \hat Q^2 r^3}{1 + \hat Q^2 r^4} \right) \d A_\perp' +
\left(\frac{\w^2}{f^2} - \frac{1}{f} \frac{k^2}{1 + \hat Q^2 r^4} \right) \d A_\perp = 0 \,,
\ee
where $\d A_\perp = \d A_x$. The equation for the longitudinal mode $\d A_\parallel = \w \d A_y + k \d A_t$ is
\be\label{eq:parallel}
\d A_\parallel'' + \frac{6 \hat Q^2 k^2 r^3 f^2 - \w^2 (1 + \hat Q^2 r^4) \left(2 \hat Q^2 r^3 f + (1 + \hat Q^2 r^4) f' \right)}{f (1 + \hat Q^2 r^4) \left(k^2 f - \w^2 (1 + \hat Q^2 r^4) \right)}\d A_\parallel'
+ \left(\frac{\w^2}{f^2} - \frac{1}{f} \frac{k^2}{1 + \hat Q^2 r^4} \right) \d A_\parallel = 0 \,.
\ee
This decoupling of charge dynamics from heat flow and momentum conservation comes at a physical cost. For instance, the translation-invariance delta function in the conductivity is not visible in the probe limit \cite{Karch:2007pd}.

We are primarily interested in zero temperature. This corresponds to setting $f = 1$ in the above equations. In this limit we recover the equations in \cite{Goykhman:2012vy}. Our objective in this paper is to understand the momentum dependence of the current spectral density in these systems at
vanishing frequency. We have managed to obtain this spectral weight analytically for the transverse modes. A low frequency expansion is implemented by setting
\be\label{eq:wexpand}
\d A_\perp = e^{i \w r} \left(\d A_\perp^{(0)} + \w \d A_\perp^{(1)} + \cdots \right) \,.
\ee
It is necessary to extract the infalling near-horizon behavior as we have done here to allow for the fact that the $\w \to 0$ and $r \to \infty$ limits do not commute. Said differently, the factor of $e^{i \w r}$ removes the irregular singular point in the equation at infinity. To zeroth order in $\w$ we then find that the solution that is regular at the horizon is
\be
\d A_\perp^{(0)}(r) = \sinh \left( \frac{k}{\sqrt{\hat Q}} \left[(-1)^{1/4} F\left(i \, \text{arcsinh} \left((-1)^{1/4} \sqrt{\hat Q} \, r \right) ,-1\right)  + \frac{\G(1/4)^2}{4 \sqrt{\pi}} \right] \right) \,.
\ee
Here $F$ is the incomplete elliptic integral of the first kind. The normalization of this solution is immaterial. The first order solution that is regular at the horizon can be written as
\be\label{eq:a1}
\d A_\perp^{(1)}(r) = c \, \d A_\perp^{(0)}(r) - i \left(r  + \frac{\sqrt{1 + \hat Q^2 r^4}}{\hat Q} \frac{d}{dr}\right)  \d A_\perp^{(0)}(r) \,.
\ee
Here $c$ is a constant. Obtaining this unexpectedly (to us at least) simple form of the first order solution required some nontrivial integration by parts of an integral expression that is obtained upon solving the differential equation for $\d A_\perp^{(1)}$. It may also be verified directly by plugging the form (\ref{eq:a1}) into the differential equation for $\d A_\perp^{(1)}$ and then using the differential equation for $\d A_\perp^{(0)}$. It is now straightforward to expand the full solution near the boundary $r \to 0$, giving $\d A_\perp = \d A_{\perp,\text{n.n.}} + \d A_{\perp,\text{n.}}  r + \cdots$. The retarded Green's function is given, as usual and up to an overall normalization, by the ratio of the normalizable and non-normalizable terms
\be
G^R_{J_\perp J_\perp}(\w,k) = \frac{\d A_{\perp,\text{n.}}(\w,k)}{\d A_{\perp,\text{n.n.}}(\w,k)} \,.
\ee
The
desired transverse spectral weight (dissipative conductivity) is then found to be
\be\label{eq:chiprobe}
\sigma_\perp(k) = \lim_{\w \to 0} \frac{\text{Im} \, G^R_{J_\perp J_\perp}(\w,k)}{\w}
= \sigma_\perp(0) \,  \left(\frac{k \mu_0}{Q} \text{csch} \frac{k \mu_0}{Q}\right)^2  \,.
\ee
Here we introduced the zero temperature limit of the chemical potential (\ref{eq:mu})
\be
\mu_0 = \lim_{T \to 0} \mu = \frac{Q}{\sqrt{\hat Q}} \frac{4 \Gamma(5/4)^2}{\sqrt{\pi}} \,.
\ee
The exact expression (\ref{eq:chiprobe}) is the main result of this section and is plotted in the following figure \ref{fig:chiperpbrane}. Note the absence of hats on $Q$ in the final expression (\ref{eq:chiprobe}). We have verified this result by also computing the spectral weight numerically.
\begin{figure}[h]
\begin{center}
\includegraphics[height=190pt]{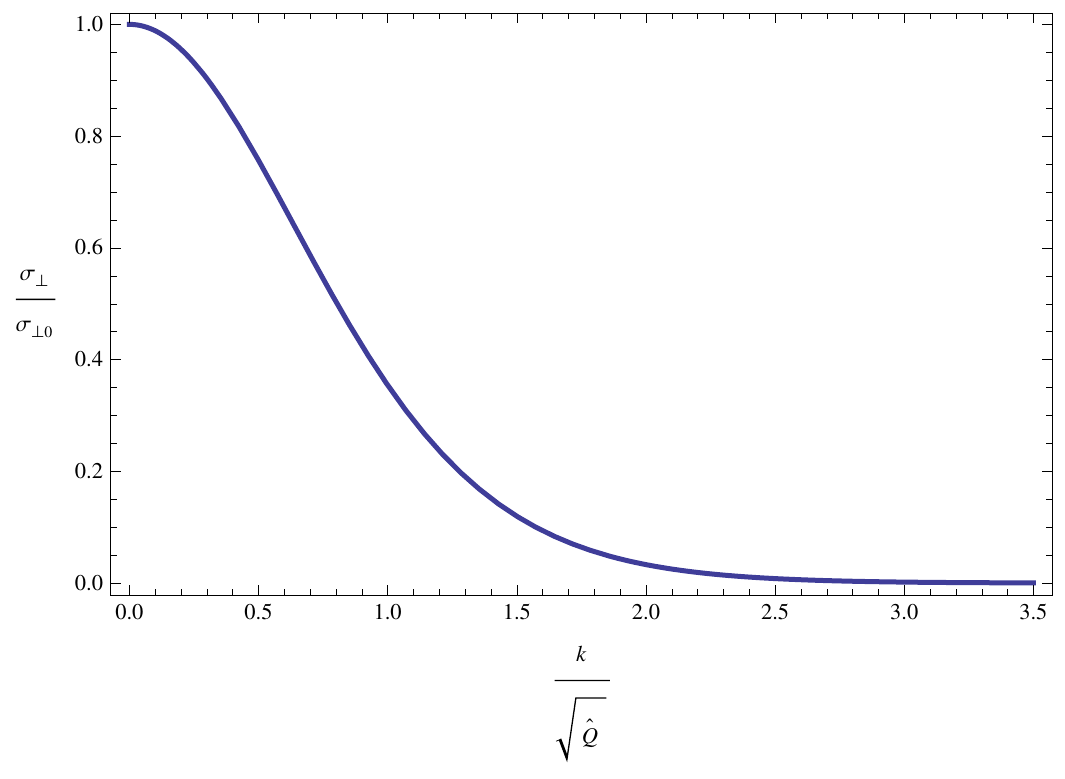}
\caption{Spectral weight (dissipative conductivity) of the transverse current at zero frequency as a function of momentum for the $D3/D5$ system at finite charge density. This is a plot of equation (\ref{eq:chiprobe}). \label{fig:chiperpbrane}}
\end{center}
\end{figure}
The main physical point to take away from (\ref{eq:chiprobe}) is that for
\be\label{eq:expboundary}
k \lesssim k_\star = \frac{Q}{2 \mu_0} \,,
\ee
the system has spectral weight at low energies. For momenta greater than this, the low energy spectral weight is exponentially suppressed.

It has already been emphasized in \cite{Hartnoll:2009ns} that gauge field fluctuations of the DBI action at finite density and zero temperature have the property that the differential equations describing the fluctuations become independent of momentum near the horizon. We can see this by taking $r \to \infty$ in (\ref{eq:perp}) and (\ref{eq:parallel}), with $f=1$ at zero temperature. It was explained that this implies the existence of low energy excitations at finite momentum. Our result (\ref{eq:chiprobe}) makes this intuition explicit and shows furthermore that these excitations are present over a precise range of momentum $0 \leq k \lesssim k_\star$.

The plot in figure \ref{fig:chiperpbrane} is satisfyingly along the lines of what one might expect for a strongly interacting quantum liquid. The well-defined vanishing of the free fermion transverse low energy spectral weight (\ref{eq:freetrans}) has been smoothened out by interactions. Nonetheless, the basic `exclusion principle' feature of low energy degrees of freedom extended in momentum space remains.

Static screening of electrical currents is determined by the real rather than the imaginary part of the transverse current Green's function. The static Green's function is also computable analytically from our solutions above for  $\d A_\perp^{(0)}$ and $\d A_\perp^{(1)}$. We obtain the transverse susceptibility
\be\label{eq:chichiperp}
\chi_\perp(k) = \lim_{\w \to 0}  G^R_{J_\perp J_\perp}(\w,k) = - |\chi_\perp(0)| \, \frac{k \mu_0}{Q} \coth \frac{k \mu_0}{Q} \,.
\ee
The magnetic response to a line of current is then obtained by Fourier transforming the
screened boundary transverse gauge field propagator $\left(k^2 - \chi_\perp(k)\right)^{-1}$.
Curiously, using the susceptibility (\ref{eq:chichiperp}) we find that all the poles of the screened propagator are at purely imaginary momenta. It follows
that the screening leads to a purely exponential falloff of the magnetic field with distance from the current, with no oscillations.
Possibly relatedly, no quantum oscillations in the magnetic susceptibility are observed in this and similar probe branes systems as a function of the magnetic field, e.g. \cite{Goykhman:2012vy}.

For the longitudinal channel and at finite temperatures, we have not been able to solve for the spectral weight and static susceptibility analytically. In particular the zeroth order longitudinal equation with $\w=0$ does not appear solvable even at zero temperature. The relative complexity of the longitudinal equation is partially due to the fact that it supports a zero sound mode \cite{Karch:2009zz, Goykhman:2012vy}. This feature dominates the spectral density at any small finite $\w$ and $k$ and is not a direct probe of structure at finite momentum. At zero temperature we solved the equations for $\d A_\parallel$ numerically in a perturbative expansion in $\w$, analogously to the expansion in (\ref{eq:wexpand}). This expansion pushes the zero sound feature down to arbitrarily low momenta. From the solution to the longitudinal equation we may obtain e.g. the retarded Green's function for density fluctuations from \cite{Goykhman:2012vy}
\be
G^R_{J^t J^t}(\w,k) = - \frac{k^2}{k^2 - \w^2} \frac{\d A_{\parallel,\text{n.}}(\w,k)}{\d A_{\parallel,\text{n.n.}}(\w,k)} \,.
\ee
The longitudinal susceptibility
\be
\chi_\parallel(k) = \lim_{\w \to 0} G^R_{J^t J^t}(\w,k) \,,
\ee
is found to have a qualitatively similar functional form to the transverse susceptibility (\ref{eq:chichiperp}), but with the opposite sign.
We have verified numerically that, similarly to the transverse case, the screened propagator, which is now $\left(k^2 + \chi_\parallel(k)\right)^{-1}$, has poles at purely imaginary momenta. It follows that point charges are exponentially screened in this theory and do not cause Friedel-type oscillations. 

\begin{figure}[h]
\begin{center}
\includegraphics[height=190pt]{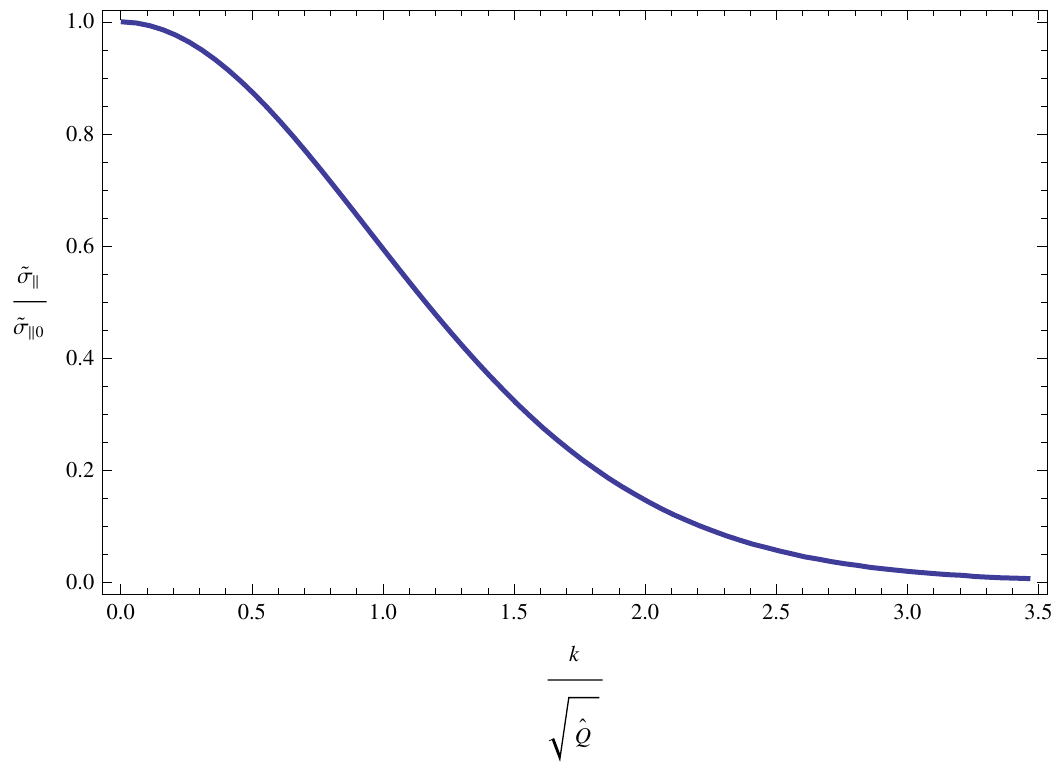}
\caption{Spectral weight (\ref{eq:w3}) of the longitudinal current at zero frequency as a function of momentum for the $D3/D5$ system at finite charge density. This quantity has been obtained numerically. The zero sound feature has been pushed to arbitrarily small momentum. \label{fig:chiparallelbrane}}
\end{center}
\end{figure}

The spectral density is found to vanish with a higher power of $\w$ than in the transverse channel. We had to work numerically to order $\d A_\parallel^{(3)}$ to extract the leading nonvanishing spectral weight. Our numerical result for
\be\label{eq:w3}
\widetilde \sigma_\parallel(k) =  \lim_{\w \to 0} \frac{k^2}{\w^2} \, \frac{\text{Im}  \, G^R_{J^t J^t}(\w,k)}{\w} \,,
\ee
is shown in figure \ref{fig:chiparallelbrane}.
Note the extra factor of $k^2/\w^2$ in the definition of $\widetilde \sigma_\parallel$ relative to the density-density spectral weight defined in (\ref{eq:weights}). That is, $\widetilde \sigma_\parallel = \sigma_\parallel(\w) k^2/\w^2$. We see therefore that at a given finite momentum there is much less low energy spectral weight in the longitudinal channel than in the transverse channel. At finite temperature, the fact that $\sigma_\parallel(\w) \sim \w^2$ will translate into the spectral weight vanishing like $T^2$ at low temperature. The above said, the spectral weight that does exist is seen in figure \ref{fig:chiparallelbrane} to be constrained to lie in a finite band of momenta. The dependence on momentum is qualitatively similar to the transverse channel shown in figure \ref{fig:chiperpbrane} and equation (\ref{eq:chiprobe}).

A quantity that can be extracted analytically at any temperature and for both channels are the momenta analogous to (\ref{eq:expboundary}), above which the spectral weight is exponentially suppressed. As the temperature is increased, we expect to see spectral weight move to higher momenta as states are Boltzmann populated. The exponential decay at large momenta can be extracted with a WKB computation, cf. \cite{Son:2002sd}. The imaginary part of the retarded Green's function is given by the probability for the excitation to tunnel from the boundary to the horizon. We can perform this computation directly at zero frequency, $\w = 0$. Mapping the equations to Schr\"odinger form and using $k \to \infty$ as the WKB parameter, one finds for both the longitudinal and transverse modes
\bea
\text{Im} \, G^R(k) & \sim & \exp \Big\{- 2 k \int_0^{r_+} \frac{dr}{\sqrt{f(r) \left(1 + \hat Q^2 r^4\right)}} \Big\} \\
& = & 
\exp \Big\{ - \frac{2 \G(5/4)}{\sqrt{\pi} \G(3/4)} \frac{k}{T} {}_2 F_1 \left(\frac{1}{4},\frac{1}{2},\frac{3}{4}, - \frac{\hat Q^2}{(\pi T)^4} \right) \Big\} \,. \label{eq:hyper}
\eea
In the low temperature limit, $T \to 0$, we recover the critical momentum in (\ref{eq:expboundary}) separating regions with and without spectral weight. The same rate  of exponential falloff with momentum is therefore seen to apply to both the transverse and longitudinal channels.
At high temperatures the spectral weight is seen to be non-negligible for momenta (cf. \cite{Son:2002sd})
\be
k <  k_\star = \frac{\sqrt{\pi} \G(3/4)}{2 \G(5/4)} T \,. \qquad (\text{as } T \to \infty)
\ee
This is precisely the anticipated Boltzmann occupation of states. The hypergeometric function in the exponent of (\ref{eq:hyper}) curtails the decrease of $k_\star$ with temperature so the finite region of momentum space in (\ref{eq:expboundary}) remains active at zero temperature.

\section{Discussion}

Our starting point has been the observation that the existence
of low energy degrees of freedom at nonzero momenta is a fundamental organizing principle
for the best characterized compressible state of matter: the Fermi liquid.
Given the central role this fact plays in the phenomenology and theory of metals,
the extent to which such momentum space structure appears in holographic phases
of matter is currently an active topic of investigation \cite{Polchinski:2012nh, Hartnoll:2012wm, Faulkner:2012gt}. In this paper we have found the first holographic identification of a strong non-analyticity in the low energy spectral weight as a function of momentum that is visible to leading order in the bulk gravity limit (\ref{eq:infzero}). In a second class of holographic theories we have found low energy spectral weight that is exponentially localized below a critical momentum (\ref{eq:chiprobe}).

Semi-local quantum criticality is among the most interesting new organizing principles to emerge
from holographic studies of finite density matter. This criticality applies to IR fixed points in which time scales but space (and hence momentum) does not. The crucial extra ingredient of semi-local criticality is that the weights with which operators scale in time are momentum-dependent. Applied to current correlators, which are the quantities that universally detect the charge dynamics in the system at leading order in the bulk gravity limit, the momentum-dependence of the scaling dimension allows a sharp momentum-space structure to emerge at zero temperature. While the local scaling itself kinematically allows low energy spectral weight at all momenta, this spectral weight only participates in charge dynamics over a finite range of momenta for which the associated current operators are relevant in the IR. This effect has an interesting resemblance to the momentum space structure of the spectral weight of a Fermi liquid.

Initial work on semi-local quantum criticality in finite density holography revolved around $AdS_2 \times \R^2$ \cite{Faulkner:2009wj}. It was emphasized in \cite{Hartnoll:2012wm} that this is a special case of holographic semi-local criticality, which is more generally parameterized by the quantity $\eta$ discussed in the introduction. For $\eta > 0$ there is no ground state entropy density. For $0 < \eta < 2$ we have found an interesting momentum space structure. We have furthermore noted that $\eta = 1$ is easily embedded into string theory and thereby realized as a finite density phase of the M2 brane CFT dual to $AdS_4 \times S^7$. The same case of $\eta = 1$ was found to exhibit mathematical simplifications and a linear in temperature resistivity in the presence of impurities.

The key question facing semi-local criticality seems to be its generality. Can it arise naturally in non-holographic systems? Is it somehow an artifact of the large $N$ limit, perhaps describing multiple Fermi surfaces that have been smeared over momentum space? Strictly local criticality is known to arise in a certain limit of spin glass systems \cite{Sachdev:2010um}. Here we are specifically interested in semi-local criticality, where there is a momentum dependence in the weights and also no entropy at zero temperature.

We suspect that the decoupling of the gauge invariant perturbations we have performed may be simpler to understand by working directly in an uplifted theory with fewer basic fields. These could be the uplifts in \cite{Cvetic:1999xp} or those in \cite{Gouteraux:2011ce}.

An interesting feature in both classes of holographic models that we considered is that there is a stronger momentum space structure in the transverse channel than the longitudinal channel. This is the opposite to free fermions. Also of interest is the fact that holography makes manifest the physical distinction between low energy spectral weight and the static susceptiblity (which controls the existence of Friedel oscillations). We have found strong momentum-space features in the spectral weight. However, the static susceptibility either does not exhibit any features -- in the probe brane case -- or is independent of IR data -- in the semi-locally critical case.
We noted that the probe brane case is also known not to exhibit quantum oscillations as function of magnetic field. It appears to be an interesting question for future work to look for quantum oscillations in the semi-locally critical cases with $0 < \eta  < 2$; it is possible that the dramatic spectral weight (\ref{eq:infzero}) will have consequences for the magnetic susceptibility. The first step is to construct the extremal dyonic black holes in these theories.

We end with a few comments on the D3/D5 theory. From a weakly coupled worldvolume perspective this theory has both charged bosons and fermions. The momentum space structure that we have nonetheless found in the strongly interacting gravity dual suggests that it is not necessary to use purely fermionic worldvolume theories to obtain strongly interacting cousins of Fermi liquids. Because we have treated the D3/D5 system in the probe limit, the Drude delta function is not visible in the conductivity \cite{Karch:2007pd}. It follows that we cannot compute the d.c. conductivity due to charged lattices or impurities in this case, as these revolve around resolving the Drude delta function \cite{Hartnoll:2012rj}. It would certainly be interesting to find a setup where the momentum space structure we have uncovered, which have the form of smoothened out Fermi liquids, could be used as an input for the study of d.c. transport.

\section{Acknowledgements}

It is a pleasure to acknowledge helpful discussions with Aristomenis Donos, Thomas Faulkner, Diego Hofman, Steve Kivelson, Hong Liu, Raghu Mahajan, Jonathan Maltz and Eva Silverstein. SAH is partially funded by an Alfred P. Sloan research fellowship and by a DOE Early Career Award. DMR is supported by a Morgridge Family Stanford Graduate Fellowship and RJA by a Stanford Humanities and Sciences Fellowship.

\appendix

\section{Free fermions}
\label{sec:free}

The propagator of a spinless nonrelativistic free fermion in Euclidean space is
\be
G(i \w_n,k) = \frac{1}{- i \w_n + \xi_k} \,,
\ee
with
\be
\xi_k = \frac{k^2}{2m} - \mu \,.
\ee
Here $m$ is the mass and $\mu$ the chemical potential (Fermi energy).
The Euclidean density-density Green's function is, not keeping track of the overall normalization,
\bea
\langle J^t J^t \rangle(i \w_n,k) & \propto & \int d\w_n' d^2k' G(i \w_n + i \w_n', k + k') G(i \w_n',k') \,,  \\
& \propto & \int d^2k' \frac{\theta(-\xi_{k'}) - \theta(- \xi_{k+k'})}{i \w_n + \xi_{k'} - \xi_{k + k'}} \,, \\
& \propto & \int_0^{k_F} q dq \int_0^{2\pi} d\theta \left(\frac{1}{i \w_n - \frac{k^2}{2m} - \frac{k q}{m} \cos\theta}
- \frac{1}{i \w_n + \frac{k^2}{2m} - \frac{k q}{m} \cos\theta} \right) \,. \label{eq:lastline}
\eea
The retarded Green's function is given by taking $i \w_n \to \w + i \epsilon$. The imaginary part is now particularly easy to compute
\be
\text{Im}\,  G_{J^t J^t}^R(\w,k) \propto  \sum_\pm \pm \int_0^{k_F} q dq \int_0^{2\pi} d\theta \d\left(\w \mp \frac{k^2}{2m} - \frac{k q}{m} \cos\theta \right)  \,.
\ee
This integral is trivial to perform and in the low energy limit we obtain
\be
\text{Im} \, \frac{G_{J^t J^t}^R(\w,k)}{\w} \; \propto \; \left\{ \begin{array}{cc}
{\displaystyle \frac{1}{k \sqrt{k_F^2 - k^2/4}}} & k < 2 k_F \\
0 & k > 2 k_F
\end{array} \right. \,,
\ee
as quoted in the introduction.

The transverse spectral weight is similarly obtained. The Euclidean Green's function is
\be
\langle J_\perp J_\perp \rangle(i \w_n,k) \; \propto \; \int d\w_n' d^2k' \left(k'^{2} - \frac{(k\cdot k')^2}{k^2} \right) G(i \w_n + i \w_n', k + k') G(i \w_n',k') \,.
\ee
Following the same manipulations as in the density channel, we obtain that the imaginary part is given by
\be
\text{Im}\,  G_{J_\perp J_\perp}^R(\w,k) \propto  \sum_\pm \pm \int_0^{k_F} q^3 dq \int_0^{2\pi} \sin^2\theta d\theta \d\left(\w \mp \frac{k^2}{2m} - \frac{k q}{m} \cos\theta \right)  \,.
\ee
These integrals are also easily performed, and give the result for the transverse channel that we quoted in the introduction.

The static susceptibilities are computed using similar methods. Taking the real instead of the imaginary part of the last line (\ref{eq:lastline}) above, after analytic continuation $i \w_n \to \w + i \epsilon$, gives the principal value of the pole rather than a delta function. These integrals can be performed to give the well known two dimensional Lindhard functions.

\section{Gauge invariant variables for semi-local models}
\label{sec:gaugeinvar}

Even though we have chosen to work in the radial gauge ($A_r = g_{\mu
  r} = 0$), there is still some gauge redundancy in each set of
variables, arising from the $U(1)$ gauge invariance and diffeomorphism
invariance. Invariance under a $U(1)$ gauge transformation
parametrized by $\lambda$ and a diffeomorphism generated by the vector
field $\xi^\mu$ tells us that the modes
\bea
  \delta \tilde{g}_{\mu\nu} &=  &\delta g_{\mu \nu} - {\cal L}_\xi
  \bar{g}_{\mu \nu} = \delta g_{\mu \nu} - (\bar{\nabla}_\mu \xi_\nu +
  \bar{\nabla}_\nu \xi_\mu) \,, \\
  \delta \tilde{A}_\mu &= & \delta A_\mu - {\cal L}_\xi \bar{A}_\mu -
  \bar{\nabla}_\mu \lambda = \delta A_\mu - (\xi^\nu \bar{\nabla}_\nu
  \bar{A}_\mu + \bar{A}_\nu \bar{\nabla}_\mu \xi^\nu) -
  \bar{\nabla}_\mu \lambda \,, \\
  \delta \tilde{\Phi} &= &\delta \Phi - {\cal L}_\xi \bar{\Phi} =
  \delta \Phi - \xi^\mu \bar{\nabla}_\mu \bar{\Phi} \,,
\eea
represent the same physical perturbation as $\{\delta g_{\mu \nu},
\delta A_\mu, \delta \Phi \}$. Here ${\cal L}_\xi X$ is the Lie
derivative of $X$ along $\xi^\mu$ and we have denoted quantities
associated with the background with bars (i.e. $\bar{\nabla}_\mu$ is
the covariant derivative associated with the background metric
$\bar{g}_{\mu \nu}$).

With Fourier transformed fluctuations $\{\delta g_{\mu \nu}, \delta
A_\mu, \delta \Phi \}$ and parameters $\{\lambda, \xi^\mu \}$ of the
form $X(t,r,x)= e^{-i(\omega t- k x)} X(r)$, the game is to find
linear combinations of the fluctuations that are invariant under the
transformations above while staying within the radial gauge and
staying within the correct parity channel. In the
transverse channel, the three modes $\{\delta A_y, \delta g_{xy},
\delta g_{yt}\}$ can be combined to form the following two gauge
invariant variables:
\be
  \psi_{\perp,1} = \delta A_y \,, \qquad \psi_{\perp,2} = r^2 ( \omega \delta g_{xy} + k
  \delta g_{yt}) \,.
\ee
In the longitudinal case a suitable set of gauge invariant modes is
\begin{align}
  \psi_{\parallel,1} &= \frac{\kappa}{e L} \partial_r \delta A_t - \frac{
    r^3}{4 } \frac{h'(r) \partial_r[ f(r) g(r)] -2 f(r) g(r)
    h''(r)}{L^2 f(r) g(r)}
  \delta g_{yy} + \frac{1}{2L^2 f(r)}  h'(r) \delta g_{tt}\\
  &= \frac{\kappa}{e L} \partial_r \delta A_t -
  \frac{\sqrt{1+\eta}}{Z_0 L^2 \eta r^{1-2/\eta} \chi(r)} \delta g_{tt} \,,
  \notag\\
  \psi_{\parallel,2} &= \delta g_{yy} + \frac{g(r)}{2 r g(r) + k^2 r^5
    g^2(r) + r^2 g'(r)} \partial_r [ r^2 (\delta g_{xx}
  + \delta g_{yy}) ]\\
  &= \delta g_{yy} + \frac{\eta \chi(r)}{2r} \frac{1}{(1+ \eta) (1+ 2
    \hat{k}^2) - (1+2\eta) \chi(r)} \partial_r\left[ r^2 \left(
      \delta g_{xx}+ \delta g_{yy} \right) \right] \,, \notag\\
  \psi_{\parallel,3} &= \frac{1}{\sqrt{1+2/\eta}} \left[\delta \Phi +
    \frac{r^3 \bar{\Phi}'(r)}{2 L^2} \delta g_{yy}\right] =
  \frac{\delta \Phi}{\sqrt{1+2/\eta}}+ \frac{r^2}{2L^2 } \delta
  g_{yy} \,,
\end{align}
where we have used the background solution (\ref{eq:fgh}) and defined
$\hat{k} = \sqrt{1+1/\eta} k/ V_0$ to simplify.

\section{Perturbation equations for semi-local models}
\label{sec:perteqns}

\subsection{Transverse modes}

The linearized Einstein-Maxwell-dilaton equations of motion for the transverse gauge invariant modes are found to be
\begin{align}
  0 ={}& \widetilde \psi_{\perp,1}''(r) + \frac{-2(1+\eta) +5\eta
    \chi(r)}{r \eta \chi(r)} \widetilde \psi_{\perp,1}'(r) + \frac{2 k
    r^{-3+2/\eta} (1+\eta)}{
    \eta [r^{4/\eta} \omega^2 - k^2\chi(r)]} \psi_{\perp,2}'(r) \\
  {}&+ \left\{ \frac{4(1+\eta)^2 [r^{4/\eta} \omega^2 - k^2
      \chi(r)]}{r^2 V_0^2 \eta^2 \chi^2(r)} - \frac{8 \omega^2
      (1+\eta)}{r^{2(1-2/\eta)} \chi(r) \left[r^{4/\eta} \omega^2 -k^2
        \chi(r) \right]}\right\} \widetilde \psi_{\perp,1}(r) \,, \notag\\
  0={}& \psi_{\perp,2}''(r) + \frac{k^2 (\eta-2) \chi^2(r) +
    r^{4/\eta} \omega^2 [\eta \chi(r) -2(1+\eta) ]}{r \eta \chi(r)
    [r^{4/\eta} \omega^2 - k^2 \chi(r)]} \psi_{\perp,2}'(r) - \frac{4
    k r^{1-2/\eta}
  }{\eta} \widetilde \psi_{\perp,1}'(r)  \\
  {}&- \frac{8 k r^{2/\eta} \omega^2 [(\eta-1) \chi(r) -(1+\eta)]
  }{\eta^2 \chi(r) [r^{4/\eta} \omega^2 - k^2 \chi(r)]} \widetilde
  \psi_{\perp,1}(r) + \frac{4(1+\eta)^2 [r^{4/\eta} \omega^2 - k^2
    \chi(r)]}{r^2 V_0^2 \eta^2 \chi^2(r)} \psi_{\perp,2}(r) \notag \,,
\end{align}
where we have redefined $\psi_{\perp,1}$ to include a trivial constant:
$\widetilde \psi_{\perp,1} = Z_0 \sqrt{1+\eta} \psi_{\perp,1}$. Taking our cue from
\cite{Edalati:2010hk}, these equations can be decoupled by introducing
\begin{align}
  \varphi_{\perp,\pm}(r) &= \frac{r^{1/2+1/\eta} \chi^{3/2}(r)}{ r^{4/\eta}
    \omega^2 - k^2 \chi(r)} \left[p_\pm(r) \widetilde \psi_{\perp,1}(r) +
    \psi_{\perp,2}'(r)\right] \,,
\end{align}
where
\begin{align}
  p_\pm(r) &= \frac{2 r^{1-2/\eta} \left[\left(r^{4/\eta} \omega^2 -
        k^2 \chi(r)\right) \left(-1\mp\sqrt{1+2 \eta \hat{k}^2}
      \right)- 2 k^2 \chi(r) \right]}{k \eta \chi(r)} \,.
\end{align}
The equations of motion for these new variables are then the decoupled
second order equations
\begin{align}
  0 &= \varphi_{t,\pm}''(r) + \left( \frac{q_\pm \left[\chi(r)
      \right]}{4 \eta^2 r^2 \chi^2(r)} + \frac{4(1+\eta)^2
      \left[r^{4/\eta} \omega^2 - k^2 \chi(r) \right] }{V_0^2 \eta^2
      r^2 \chi^2(r) } \right) \varphi_{t,\pm}(r) \, ,
\end{align}
where $q_\pm[\chi(r)]$ is given by:
\begin{align}                                                                    q_\pm \left[\chi(r) \right] &= \left[ 4 (1+\eta)^2 - 8(1+\eta)
    \left(3 - \eta \pm 2\sqrt{1 + 2 \eta \hat{k}^2} \right)\chi(r) -
    15\eta^2 \chi^2(r) \right]\,.
\end{align}

\subsection{Longitudinal modes}

The linearized Einstein-Maxwell-dilaton equations of motion for the gauge-invariant longitudinal modes are found to take the form
\begin{align}
  0 &= \psi_{\parallel,i}''(r) + \alpha_{ij}(r) \psi_{\parallel,j}'(r) +
  \beta_{ij}(r) \psi_{\parallel,j}(r) + \frac{4 (1+\eta)^2 [r^{4/\eta}
    \omega^2 -k^2 \chi(r)]}{V_0^2 \eta^2 \chi^2(r)} \psi_{\parallel,i}(r) \,,
\end{align}
where $\alpha_{12} = \alpha_{21} = \alpha_{31} = \alpha_{32} =
\beta_{12} = \beta_{32} = \beta_{33} = 0$, 
\begin{align*}
  \alpha_{11} ={}& \frac{-2(1+\eta) + (4+7\eta) \chi(r)}{r \eta
    \chi(r)} \,, \qquad \alpha_{33} = \frac{1}{r} - \frac{2 (1+\eta)}{
    r\eta \chi(r)}  \,, \\
  \alpha_{22} ={}& \frac{5}{r} - \frac{2(1+\eta)}{r \eta \chi(r)
    \gamma(r)} \Big\{(1+\eta)^2 (1+ 2\hat{k}^2) - (1+ \eta) (4 + 5
  \eta) (1 + 2 \hat{k}^2) \chi(r) \\
  &\qquad \qquad \qquad \quad\; + \left[3 + 5\eta + 4 \eta^2 + 4(1+\eta)
    (2+\eta) \hat{k}^2\right] \chi^2(r) \Big\} \,, \\
  \alpha_{23} ={}& \frac{8(1+\eta)^2 (2+\eta) (1+ 2 \hat{k}^2)
    \left(\chi(r) - 1\right)}{r^3 \eta \gamma(r)} \,,
\end{align*}
and 
\begin{align*}
  \beta_{11} ={}& \frac{-2(1+\eta)[12+\eta(16+3\eta)] + (2+\eta)
    (2+3\eta)^2 \chi(r)}{r^2 \eta^2 (2+\eta) \chi(r)} \,, \qquad
  \beta_{13} ={} \frac{-16 \hat{k}^2 r^{-5-2/\eta} (1+\eta)^2}{
    \eta^2 \chi(r)} \,, \\
  \beta_{21} ={}& \frac{-4 (1+\eta)}{r^{1-2/\eta}\eta \chi(r)
    \gamma(r) } \left\{(1+\eta) (4 \hat{k}^4 - 1) -2 \eta (1+2
    \hat{k}^2) \chi(r) + 3(\eta-1)\chi^2(r) \right\} \,, \\
  \beta_{22} ={}& \frac{4}{r^2 \eta \chi(r) \gamma(r)}
  \Big\{2(1+\eta)^3 (2 \hat{k}^4 -\hat{k}^2 -1)- 2 (1+\eta)^2 (1+ 2
  \hat{k}^2) \left[ (2+\eta) \hat{k}^2 -3 (1+\eta) \right]\chi(r)  \\
  &\qquad \qquad \quad \;\;\; - (1+\eta) \left[4 + 5\eta + 6 \eta^2+2
    (5 + 6\eta + 4 \eta^2) \hat{k}^2\right] \chi^2(r) \\
  &\qquad \qquad \quad \;\;\;+ \eta (\eta-1)(1+2\eta)
  \chi^3(r) \Big\} \,, \\
  \beta_{23} ={}& \frac{-8(1+\eta)}{r^4 \eta \chi(r) \gamma(r)}
  \Big\{(1+\eta)^2(4 \hat{k}^2-1) -2 (1+\eta) (1+2
  \hat{k}^2) [(2+\eta) \hat{k}^2-1] \chi(r)  \\
  &\qquad \qquad \quad \;\;\;+ (\eta-1) [1+\eta + 2 (2+\eta)
  \hat{k}^2]  \chi^2(r) \Big\} \,, \\
  \beta_{31} ={}& \frac{-4 r^{1+2/\eta} (1+\eta)}{\eta(2+\eta)
    \chi(r)} \,,
\end{align*}
with $\hat{k} = \sqrt{1+1/\eta} \, k/V_0$ and $\gamma(r)$ given by
\begin{align}
  \gamma(r) &= \left[(1+\eta) (1+2 \hat{k}^2) - (\eta-1) \chi(r)
  \right] \left[(1+\eta) (1+ 2 \hat{k}^2) - (1+2\eta)\chi(r)\right] \,.
\end{align}

To clean this up, we first get rid of the first derivative terms by
writing the modes as $\psi_{\parallel, i} = \zeta_i(r)
\tilde{\psi}_{\parallel,i}$, where the $\zeta_i(r)$ are given by
\begin{align*}
  \zeta_1(r) &= \frac{r^{-5/2-1/\eta}}{\sqrt{\chi(r)}} \,,&
  \zeta_2(r) &= \frac{r^{-3/2+1/\eta}}{\sqrt{\chi(r)}}
  \frac{(1+\eta)(1+2 \hat{k}^2)-(\eta-1)\chi(r)}{(1+\eta)(1+2
    \hat{k}^2) - (1+ 2\eta)\chi(r)}  \,,& 
  \zeta_3(r) &= \frac{r^{1/2+1/\eta}}{\sqrt{\chi(r)}} \,.
\end{align*}
With these definitions, the modes $\tilde{\psi}_{\parallel,1/3}$
satisfy equations of the form $0=\tilde{\psi}_{\parallel,i}'' +
\tilde{\beta}_{ij} \tilde{\psi}_{\parallel,j}$, with
$\tilde{\beta}_{12}= \tilde{\beta}_{32} = 0$. The equation for
$\tilde{\psi}_{\parallel,2}$ is not of this form, but if we look at
the linear combination $\varphi_{\parallel,2} =
\tilde{\psi}_{\parallel,2} + p(r) \tilde{\psi}_{\parallel,3}$, where
\begin{align}
  p(r) &= \frac{\chi(r) - 1}{1+(1+\eta)
    \hat{k}^2}\frac{(1+\eta)(2+\eta)(1+2\hat{k}^2)}{(1+\eta)
      (1+ 2\hat{k}^2) - (\eta-1) \chi(r)} \,,
\end{align}
then $\varphi_{\parallel,2}$ satisfies an equation of the form
$0=\varphi_{\parallel,i}'' + \tilde{\beta}_{ij} \varphi_{\parallel,j}$.  Note
that at $T=0$, $\chi(r) = 1$ so $p(r)=0$ and this combination is
unneccesary to decouple the fields (e.g. $\alpha_{23} = 0$ at
$T=0$). Finally, we clean up notation a bit by letting
$\varphi_{\parallel,i} = \tilde{\psi}_{\parallel,i}$ for $i=1,3$ with
$\varphi_{\parallel,2}$ defined as above, so the equations of motion for
the $\varphi_{\parallel,i}$ are
\begin{gather*}
  0 = \varphi_{\parallel,i}''(r) + \tilde{\beta}_{ij}(r)
  \varphi_{\parallel,j}(r) + \frac{4 (1+\eta)^2 \left[r^{4/\eta} \omega^2 -
      k^2 \chi(r) \right]}{r^2 \eta^2 V_0^2 \chi^2(r)}
  \varphi_{\parallel,i}(r) \,,
\end{gather*}
where $\tilde{\beta}_{12}=\tilde{\beta}_{32}=0$ -- this means that there are self-contained equations for $\varphi_{\parallel,1}$ and $\varphi_{\parallel,3}$ that do not involve $\varphi_{\parallel,2}$ with
\begin{align*}
  \tilde{\beta}_{11} ={}& \frac{1}{4r^2} - \frac{2(1+\eta)^2
    (10+\eta)}{r^2 \eta^2 (2+\eta) \chi(r)} + \frac{(1+\eta)^2}{r^2
    \eta^2 \chi^2(r)} \,,& \tilde{\beta}_{13} ={}& \frac{-16
    (1+\eta)^2\hat{k}^2}{r^2 \eta^2 \chi(r)} \,, \\
  \tilde{\beta}_{31} ={}&  \frac{-4(1+\eta)}{r^2 \eta(2+\eta) \chi(r)}
  \,,& \tilde{\beta}_{33} ={}& \frac{1}{4r^2} - \frac{2
    (1+\eta)^2}{r^2 \eta^2 \chi(r)} + \frac{(1+ \eta )^2}{r^2 \eta^2
    \chi^2(r)} \,,
\end{align*}
-- while $\varphi_{\parallel,2}$ couples to both
$\varphi_{\parallel,1}$ and $\varphi_{\parallel,3}$:
\begin{align*}
  \tilde{\beta}_{21} ={} \frac{-4(1+\eta)}{r^2 \eta \chi(r)
    \tilde{\gamma}(r)} \Big\{& (1+\eta) (1+ 2\hat{k}^2)
  \left[2(1+\eta)
    \hat{k}^4 - (1+3 \eta) \hat{k}^2 - 2 - \eta \right]  \\
  &+2 (1+ 2\hat{k}^2) \left[ \eta^2 + (1+\eta) \hat{k}^2 \right]
  \chi(r) + (\eta-1) \left[ 2-\eta +
    (1+\eta) \hat{k}^2\right] \chi^2(r) \Big\} \,, \\
  \tilde{\beta}_{22} ={} \frac{1+(1+\eta) \hat{k}^2}{4 r^2 \eta^2
    \chi^2(r) \tilde{\gamma}(r)} \Big\{&4 (1+\eta)^4 (1+2 \hat{k}^2)^2
  + 16 (1+\eta)^3 (1+ 2\hat{k}^2) \left[2-\eta + (3\eta-1) \hat{k}^2
  \right] \chi(r)\\
  &- (1+\eta)^2 \left[76 + 160 \hat{k}^2 + 8 \eta (1 + 4 \hat{k}^2)^2
    + \eta^2 (124 \hat{k}^4 + 28 \hat{k}^2 - 21) \right]
  \chi^2(r) \\
  &-2 (\eta^2-1) \left[12 + 16\eta + 5\eta^2 + 2\left(16 +
      \eta(16+\eta)\right) \hat{k}^2 \right]
  \chi^3(r) \\
  &+ \eta^2 (\eta-1)^2 \chi^4(r) \Big\} \,, \\
  \tilde{\beta}_{23} ={} \frac{-8(1+\eta)}{r^2 \eta^2 \chi(r)
    \tilde{\gamma}(r) }\Big\{ &(1+\eta)^2 (1 + 2\hat{k}^2) \left[ 2
    \eta(1+\eta) \hat{k}^4-2(\eta^2-1)\hat{k}^2 - \eta  \right] \\
  &- 2(1+\eta)(1+ 2 \hat{k}^2) \left[ \eta(1+\eta) (2+\eta) \hat{k}^4
  \right.\\
  &\qquad \qquad \qquad \qquad \quad \left.+\left(2(1+\eta)-\eta^2
      (2 + \eta) \right) \hat{k}^2 - \eta \right] \chi(r) \\
  &- (\eta-1) \left[ 2 (1+\eta) (2+\eta) \hat{k}^4 + 2 \hat{k}^2
    -\eta(\eta+1) \right] \chi^2(r) \Big\} \,.
\end{align*}
Here $\tilde{\gamma}(r)$ is given by
\begin{align}
  \tilde{\gamma}(r) &= \left[1+(1+\eta) \hat{k}^2 \right]
  \left[(1+\eta) (1+2 \hat{k}^2) - (\eta-1) \chi(r) \right]^2 \,.
\end{align}
The $T=0$ case simplifies considerably, and we can fully decouple
these equations; we do this in the main text. For $T \neq 0$,
$\varphi_{\parallel,1}$ and $\varphi_{\parallel,3}$ are easily decoupled.
In general, it is not clear whether a decoupled equation can be found for
$\varphi_{\parallel,2}$, except in
the case $\eta=1$ for which simplifications occur, as we discuss in the main text.
Even if we were able to decouple the equations 
for $T\neq 0$ and $\eta \neq 1$, we don't expect to be able to solve them
analytically.

\end{document}